\documentclass[prb,twocolumn,longbibliography,amsmath,superscriptaddress,amssymb,nobibnotes]{revtex4-2}

\usepackage{graphicx}
\usepackage{dcolumn}
\usepackage{bm}
\usepackage{amssymb}
\usepackage{amsmath}
\usepackage{color}
\usepackage[dvipsnames]{xcolor}
\usepackage{placeins}
\usepackage{epstopdf}
\usepackage{multirow}
\usepackage[normalem]{ulem}
\usepackage[linkcolor=blue,urlcolor=blue,citecolor=blue,colorlinks=true]{hyperref}

\begin{document}

\title{Competing magnetic order in EuPd$_3$Si$_2$}

\author{Michelle Ocker}
\affiliation{
 Kristall- und Materiallabor, Physikalisches Institut, 
Goethe-Universit\"at Frankfurt, 60438 Frankfurt/M, Germany
}
\author{Franziska Walther}
\affiliation{
 Kristall- und Materiallabor, Physikalisches Institut, 
Goethe-Universit\"at Frankfurt, 60438 Frankfurt/M, Germany
}
\author{Nour Maraytta}
\affiliation{Institute for Quantum Materials and Technologies, Karlsruhe Institute of Technology, Kaiserstr. 12, 76131 Karlsruhe, Germany}
\author{Matthieu Le Tacon}
\affiliation{Institute for Quantum Materials and Technologies, Karlsruhe Institute of Technology, Kaiserstr. 12, 76131 Karlsruhe, Germany}
\author{Michael Merz}
\email{michael.merz@kit.edu}
\affiliation{Institute for Quantum Materials and Technologies, Karlsruhe Institute of Technology, Kaiserstr. 12, 76131 Karlsruhe, Germany}
\affiliation{Karlsruhe Nano Micro Facility (KNMFi), Karlsruhe Institute of Technology, Kaiserstr. 12, 76131 Karlsruhe, Germany}

\author{Cornelius Krellner}
\affiliation{
 Kristall- und Materiallabor, Physikalisches Institut, 
Goethe-Universit\"at Frankfurt, 60438 Frankfurt/M, Germany
}
\author{Kristin Kliemt}
\email{kliemt@physik.uni-frankfurt.de}
\affiliation{
 Kristall- und Materiallabor, Physikalisches Institut, 
Goethe-Universit\"at Frankfurt, 60438 Frankfurt/M, Germany
}

\date{\today}
\begin{abstract}
\noindent
\newline
Single crystals of EuPd$_3$Si$_2$ were grown using a high-temperature EuPd-flux method. The material was structurally and chemically characterized by single-crystal x-ray diffraction, powder x-ray diffraction, Laue method and energy-dispersive x-ray spectroscopy.
The structural analysis confirmed the orthorhombic crystal structure (space group $Imma$) but revealed differences in the lattice parameters and bond distances in comparison to \cite{sharma2023crystal}. \\
The composition is close to the ideal 1:3:2 stoichiometry with an  occupation of 7 \% of the Si sites by Pd. The heat capacity, electrical resistivity, and magnetic susceptibility show two magnetic transitions indicating magnetic ordering below $T_{\rm N1}=$ 61\,K and a spin reorientation at $T_{\rm N2}=$ 40\,K. The orthorhombic material shows magnetic anisotropy with field applied along the three main symmetry axes, which is summarized in the temperature-field phase diagrams. The susceptibility data hint to an alignment of the magnetic moments along $[100]$ between $T_{\rm N1}$ and $T_{\rm N2}$. Below $T_{\rm N2}$ the magnetic structure changes to an arrangement with moments canted away from $[100]$.  In contrast to published work \cite{sharma2023crystal}, the single crystals investigated in this study are suggested to show antiferromagnetic (AFM) order below $T_{\rm N1}$ instead of ferromagnetism that sets in at higher $T_{\rm C1}=78\,\rm K$ which might originate from certain differences in the structure, composition or defects that have an impact on the dominant coupling constants of the RKKY interaction. 
\end{abstract}

\maketitle

\section{Introduction}\noindent\\
During the past decades, the study of Eu-based compounds has garnered significant attention due to the intriguing magnetic properties imparted by the 4$f$ electrons of Eu which are highly localized and contribute to a rich variety of magnetic phenomena, including valence transitions \cite{sampathkumaran1981new, seiro2013complex, honda2017pressure, Onuki2017}, colossal magnetoresistance \cite{krebber2023,manna2014}, quantum criticality \cite{repvcek2024multiferroic}, complex forms of magnetic order \cite{ hossain2001complex, jiang2005complex}, and even skyrmion lattices were found \cite{kaneko2018}.
In most compounds, Eu is in a Eu$^{2+}$ state ($4f^7$ electronic configuration: S = 7/2, L = 0, J = 7/2) that orders magnetically \cite{seiro2013complex}, while there are a few compounds where Eu is in the Eu$^{3+}$ state ($4f^6$: S = L = 3, J = 0) which are non-magnetic \cite{seiro2013anomalous} and some that are intermediate valent \cite{sampathkumaran1981new, seiro2011stable}. 
EuPd$_3$Si$_2$ crystallizes in a (pseudo hexagonal) orthorhombic ErRh$_3$Si$_2$ structure type \cite{Cenzual1988} with the space group $Imma$. This structure deviates from the higher symmetric hexagonal CaCu$_5$ structure type (space group $P6/mmm$) by a small distortion in the CaCu$_5$ $a-b$ plane and a doubling of the $c$ lattice parameter. 
In EuPd$_3$Si$_2$, Eu is divalent with J=S=7/2 and the magnetism is governed by the Ruderman-Kittel-Kasuya-Yosida (RKKY) interaction \cite{sharma2023crystal}. According to previous neutron diffraction results, the material shows ferromagnetic (FM) order of Eu$^{2+}$ below $T_{\rm C1}$ = 78\,K and at $T_{\rm C2}$ = 5\,K a spin reorientation occurs towards a phase with moments aligned along $[100]$ at $1.6\,\rm K$ \cite{sharma2023crystal}.
Deviations from highly symmetrical arrangements, such as distortions or non-centrosymmetrical environments, can lead to magnetic anisotropies that become visible when external magnetic fields are applied.
In the case of EuPd$_3$Si$_2$ the occurrence of different magnetic phases could be caused by anisotropic magnetic exchange interactions given by the orthorhombic crystal structure with the four different distances between Eu ions as illustrated in Fig.~\ref{fig:struc}.
The oscillatory character of the RKKY interaction depends on interatomic distances, e.g.
the distance between neighboring magnetic ions can influence both the type of magnetic order (e.g. FM or AFM) and the magnitude of the magnetic interactions \cite{paschen2001structural,nikitin2002influence,jiang2005complex}.  
\\
In Ref.~\cite{sharma2023crystal}, the magnetic structure of EuPd$_3$Si$_2$ was studied by neutron powder diffraction, which revealed FM order with moments aligned along $[100]$ at $1.6\,\rm K$. However, magnetization data suggests weak AFM correlations and a possible canting above $T_{\rm C2}$. Due to the high neutron absorption by Eu, the quality of the neutron powder diffraction data was limited, and the authors mention that no differences in the magnetic structure were found below and above $T_{\rm C2}$. This motivated the present work where a detailed study of the magnetic properties of EuPd$_3$Si$_2$ via magnetization measurements along the different crystallographic directions is presented.
Our data indicate that even subtle variations in the lattice parameters can significantly affect the magnetic structure of this material. Based on the magnetization data, it is suggested that the samples grown in this study exhibit an AFM order below $T_{\rm N1} = 61$\,K with the moments aligned along the Eu chains running parallel to the [100] direction. Upon further cooling, a magnetic reorientation occurs below $T_{\rm N2} = 40$\,K, where the moments become slightly canted away from [100], in contrast to the results reported in Ref.~\cite{sharma2023crystal}.

\section{Experimental}\noindent
Single crystals were grown in a box furnace (Therm Concept) using  Pd (99.99\%, rod, Heraeus), Si (99.9999\%, pieces, Cerac) and Eu (99.99\%, chunks,
EvoChem).
The crystal structure is probed by both single-crystal and powder x-ray diffraction (XRD) methods. Single-crystal XRD measurements were performed on a high-flux, high-resolution, rotating anode Rigaku Synergy-DW (Mo/Ag) diffractometer using Mo $K_\mathrm{\alpha}$ radiation ($\lambda$ = 0.7107 {\AA}). The system is equipped with pairs of precisely manufactured Montel mirror optics, a motorized divergence slit which was set to 5 mrad for these measurements, and a background-less Hypix-Arc150$^{\circ}$ detector which guarantees the lowest reflection profile distortion and ensures that all reflections are detected under equivalent conditions.
The specimens had a size of $\approx$ 30 $\times$ 30 $\times$ 20 $\mu$m$^3$ and were measured to a resolution better than 0.5 {\AA} and exhibited no mosaic spread and no additional reflections from secondary phases, highlighting their high quality and allowing for excellent evaluation using the latest version of the CrysAlisPro software package \cite{CrysAlis}.  The crystal structure of the compound was refined using JANA2006 \cite{Vaclav_229_2014},  including all averaged symmetry-independent reflections (I $>$ 2 $\sigma$).  Unit cell and space group were determined, atoms were localized within the unit cell using random phases and the structure was completed and solved using difference Fourier analysis. The structural refinements converged well, exhibiting excellent reliability factors (see Tables SI and SII in the supplemental material \cite{supplementalinfo_EuPd3Si2_2025} for residuals $wR_\mathrm{2}$,  $R_\mathrm{1}$,  and goodness of fit, GOF,  values).   
Further structural analysis was performed on powdered single crystals by powder x-ray diffraction (PXRD) using a Bruker D8 diffractometer in Bragg-Brentano geometry with Cu K$\alpha$ radiation ($\lambda$ = 1.5406\,\AA). Energy-dispersive x-ray spectroscopy (EDX) was used to investigate the chemical composition of the crystals. The orientation of the samples was determined using a Laue device with white x-rays from a tungsten anode. 
In preparation of physical measurements along the main symmetry directions, the samples were oriented and cut to produce planes perpendicular to the crystallographic directions $[100]$, $[010]$, and $[001]$. 
Heat capacity, magnetic susceptibility, and resistivity measurements were performed using the commercial measurement options of a Quantum Design Physical Property Measurement System (PPMS).
\section{Results}\noindent 
\subsection{Crystal growth}
\noindent
Single crystals of EuPd$_3$Si$_2$ were obtained during attempts to grow the related compound EuPd$_2$Si$_2$ from EuPd-flux with an initial melt stoichiometry of Eu~:~Pd~:~Si~=~1.45~:~2.5~:~1.5. 
Previously, a Pd-Si prereaction was helpful in reducing the high melting points of Pd and Si \cite{Kliemt2022a}. Therefore, Pd and Si were prereacted in an argon arc furnace to form the binary compound PdSi ($T_M\approx900 ^{\circ}\rm C$) and placed in an inner graphite crucible together with Eu. Ta or Al$_2$O$_3$ were not used as crucible materials, because both materials have been shown to be attacked by the melt \cite{Kliemt2022a}, leading to a potential unintended doping of the crystals.
\begin{figure*}
    \centering
    \includegraphics[width=1\linewidth]{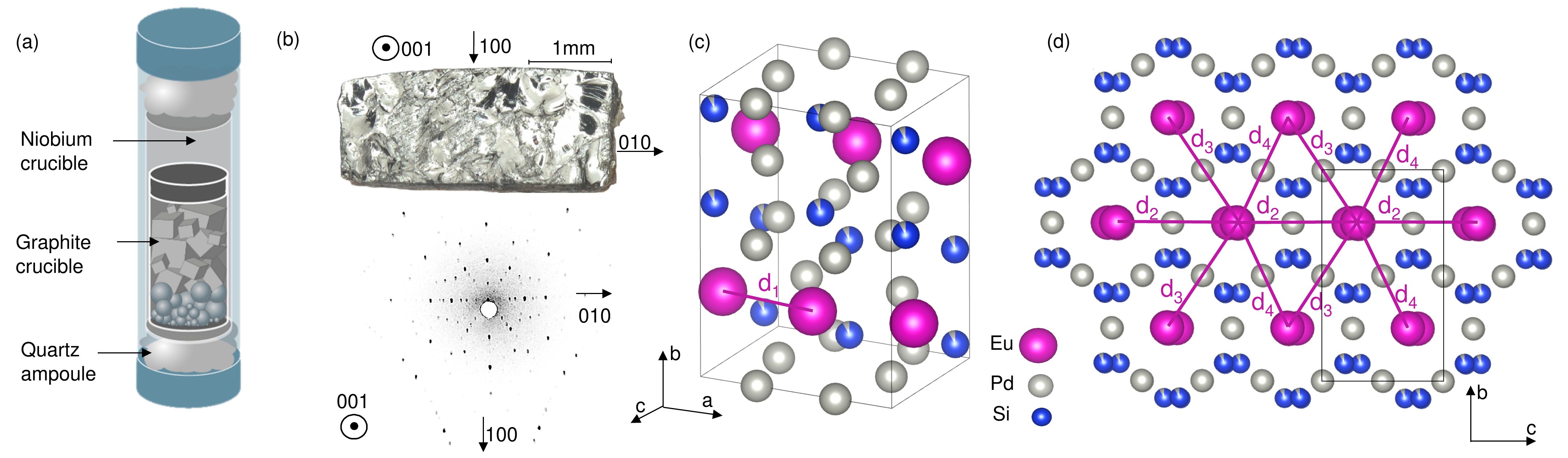}
    \caption{(a) Schematic drawing of the set up  consisting of a graphite crucible with the elements in a sealed  niobium crucible, that was welded under vacuum in a quartz ampoule. (b) A sample cut along the main symmetry directions with the corresponding Laue image.  (c) Unit cell of EuPd$_3$Si$_2$ with Pd excess. The distance between the Eu-atoms along the $a$ axis is d$_{\rm 1}$~=~3.6258(1)\,\AA.  (d) Unit cell projected onto the $b-c$ plane. The distances between the six surrounding Eu-atoms d$_{\rm 2}$~=~5.7393(5)\,\AA, d$_{\rm 3}$~=~6.0062(3)\,\AA, and d$_{\rm 4}$~=~5.5800(3)\,\AA \:are depicted.}
    \label{fig:struc}
\end{figure*}
As depicted in Fig.~\ref{fig:struc}(a), the graphite crucible was first enclosed in a sealed niobium crucible under an argon atmosphere. To prevent oxidation of niobium, it was sealed in a quartz ampoule. The ampoule was placed in a box furnace and heated fast with a rate of 100 $^{\circ}$C/h to 800\,$^\circ$C, as Eu is still in the solid state below this temperature. Subsequently, the temperature was gradually increased, with a rate of 50$^{\circ}$C/h, to 1230\,$^\circ$C to avoid rapid evaporation of Eu, and the materials were homogenized for 2 hours at this temperature. The ampoule was then slowly cooled with a rate of 2\,$^{\circ}$C/h to 1000\,$^\circ$C and fast cooled to room temperature. After the growth experiments, the graphite crucible was not attacked by the melt. A typical example of an extracted crystal is shown in Fig.~\ref{fig:struc}(b).\\
The crystal growth of the compound EuPd$_3$Si$_2$ as a by-product during the growth of the compound EuPd$_2$Si$_2$ in a Bridgman setup was previously reported in Ref.~\cite{sharma2023crystal}.
There, the authors describe that Eu, Pd, Si were prereacted in a 1:2:2 ratio in an arc furnace and heated to $T_{\rm max}=1350~^{\circ}$C in a tantalum tube in the Bridgman furnace. In addition to the two resulting compounds EuPd$_2$Si$_2$ and EuPd$_3$Si$_2$, the compound TaSi$_2$ and other phases formed \cite{sharma2023crystal}. 

\subsection{Structural and chemical analysis}\noindent
Single-crystal x-ray diffraction (SC-XRD) measurements at room temperature (RT) were refined in two candidate space groups (SGs): centrosymmetric $Imma$,  as suggested in Ref.~\cite{sharma2023crystal}, and its non-centrosymmetric subgroup $Imm2$.  Refinement results are summarized in Tab. SI and SII in the supplemental material \cite{supplementalinfo_EuPd3Si2_2025}.  Both models yield very similar structures and almost identical bond lengths, despite $Imm2$ involves splitting of certain atomic sites into additional Wyckoff positions. Displacement parameters were refined anisotropically, with equivalent values $U_{\rm eq}$ shown in Tab. SI and SII in the supplemetal material \cite{supplementalinfo_EuPd3Si2_2025}.  Statistical errors are derived from the refinements. Convergence was very good for both SGs, with excellent reliability factors ($wR_\mathrm{2}$,  $R_\mathrm{1}$,  and GOF).  No indications of twinning originating from a hexagonal-to-orthorhombic phase transition at higher temperatures---manifested by a threefold reflection splitting expected for such a transition in the precession images of the reciprocal lattice reconstructed from the collected SC-XRD data---were observed. \\
$Imma$ and $Imm2$ provide absolutely comparable agreement factors, however, $Imma$ requires fewer positional and displacement parameters for describing the structure.  Importantly, reducing the symmetry to $Imm2$ removes inversion symmetry, necessitating twinning to recover it, thereby producing two merohedral inversion twins.  Therefore, the Flack parameter $f$ was introduced to determine the absolute configuration of the non-centrosymmetric model, refining the relative fractions of the two inversion twins. For $Imm2$, $f$ refines to $\approx$ 50 \%, consistent with perfect twinning or retention of centrosymmetry. The $Imm2$ model which has a polar axis along the $c$ direction and, in principle, even has the potential for pyroelectricity offers no structural advantage, increases the number of refined parameters, and does not lead to improved agreement factors.\\
By taking all of the above mentioned factors into account and in the absence of independent evidence for acentricity, the higher symmetry $Imma$ is conventionally preferred.  Consistent with the structure published in Ref.~\cite{sharma2023crystal}, the subsequent discussion will therefore predominantly focus on the $Imma$ solution although SG $Imm2$ cannot be completely ruled out. 
\begin{table}[b]
    \centering
    \begin{tabular}{|c|c|c|c|}
    \hline
         a [\AA]&b[\AA]&c[\AA]&Reference \\
         \hline
         \hline
         7.1483(4)& 10.0743(4)  & 5.7506(2)  & SC XRD \cite{sharma2023crystal}\\
         7.1314(2)& 10.0544(2)  & 5.7325(9)  & PXRD \cite{sharma2023crystal}\\
         7.2003(1)  & 10.0370(2)  & 5.7393(1) & SC XRD, this work\\
         7.1948(5) & 10.0319(4) & 5.7352(5)& PXRD, this work\\
         \hline
    \end{tabular}
    \caption{Comparison of the lattice parameter $a, b$ and $c$ of EuPd$_3$Si$_2$ obtained  through SC-XRD and PXRD and those reported in Ref.~\cite{sharma2023crystal}.}
    \label{tab:lattice}
\end{table}
The results from our SC-XRD were used as input for the PXRD data, and as can be seen (Fig.~S1 in Ref.~\cite{supplementalinfo_EuPd3Si2_2025}),  the powder data are fully consistent with SG $Imma$ as well.
As summarized in Tab.  \ref{tab:lattice},  our SC-XRD and PXRD lattice parameters agree closely with each other but differ noticeably from those in Ref.  \cite{sharma2023crystal},  with the $a$ lattice parameter being $\approx$ 0.8 \% larger and the $b$ lattice parameter $\approx$ 0.6 \% smaller. These are significant changes, already indicative of structural modifications within the unit cell. It should be emphasized that the PXRD powder samples were prepared by grinding large single crystals. Consequently, the PXRD measurements reflect the properties of a macroscopic sample, in contrast to the microscopically small crystals employed for SC-XRD, which may not be fully representative of the bulk magnetization and transport measurements.\\
Furthermore, the current SC-XRD refinements reveal that 7 \%  Pd are substituted on the Si site. Such replacement effects potentially affect the corresponding crystal fields around the Eu, Si, and Pd atoms. Therefore, with reference to Fig.~\ref{fig:struc}(c) and (d),  especially the Eu–Eu bond distances are examined in greater detail below, since in EuPd$_3$Si$_2$ the Eu atoms are mainly responsible for the magnetic properties. \\
As can be seen from Fig.~\ref{fig:struc}(d),  the Eu atoms adopt a pseudo-hexagonal arrangement, with the distances d$_{\rm 2}$~=~5.7393(5)\,\AA, d$_{\rm 3}$~=~6.0062(3)\,\AA, and d$_{\rm 4}$ = 5.5800(3)\,{\AA} between the Eu atoms within the $b$–$c$ plane, and perpendicular to it the pseudo-hexagonal layers are strongly connected by the distance d$_{\rm 1}$~=~3.6258(1) {\AA} along the $a$ direction (Fig.~\ref{fig:struc}(c)). For completeness, the values in the case of the non-centrosymmetric SG $Imm2$ would be:  d$_{\rm 2'}$ = 5.7394(5) {\AA},  d$_{\rm 3'}$ = 6.0064(3) {\AA},  d$_{\rm 4'}$ = 5.5799(3) {\AA},  and d$_{\rm 1'}$ = 3.6258(1) {\AA}.  In principle, the two SGs lead to identical results. For comparison, the corresponding values from Ref.  \cite{sharma2023crystal} are:  d$_{\rm 2*}$ =  5.7506 {\AA},  d$_{\rm 3*}$ = 6.0072 {\AA},  d$_{\rm 4*}$ = 5.6134 {\AA},  and d$_{\rm 1*}$ =  3.5962 {\AA}. In other words, all bond lengths within the $b$–$c$ plane are slightly shorter in our current data, while those along the $a$-direction are slightly longer. Similar deviations between the current and the published data are also found for the Si and Pd coordination shells. These effects can be attributed to the presence of 7 \% Pd on the Si site in the current samples. As we will see later, it is most probably the changes in the Eu–Eu distances that can significantly modify the magnetic behavior of the samples via RKKY interaction. \\
The chemical analysis using EDX (given with the statistical deviation of several measurements on one sample) shows that the elements are present in a ratio of Eu:Pd:Si = (15.7$\pm$0.7)~:~(52.3$\pm$0.8)~:~(32.0$\pm$0.3). Adding a systematic error of measurement of at least 2~at.\%, the composition is close to the ideal composition of Eu~:~Pd~:~Si = 16.67~:~50~:~33.33. A Pd excess of 7\% as detected by SC XRD is only marginally above our resolution limit of EDX and is hardly detectable. 
The results of the chemical analysis are therefore consistent with those of the structural characterization.
\subsection{Heat capacity}
\begin{figure}[ht!]
    \centering
    \includegraphics[width=0.47\textwidth]{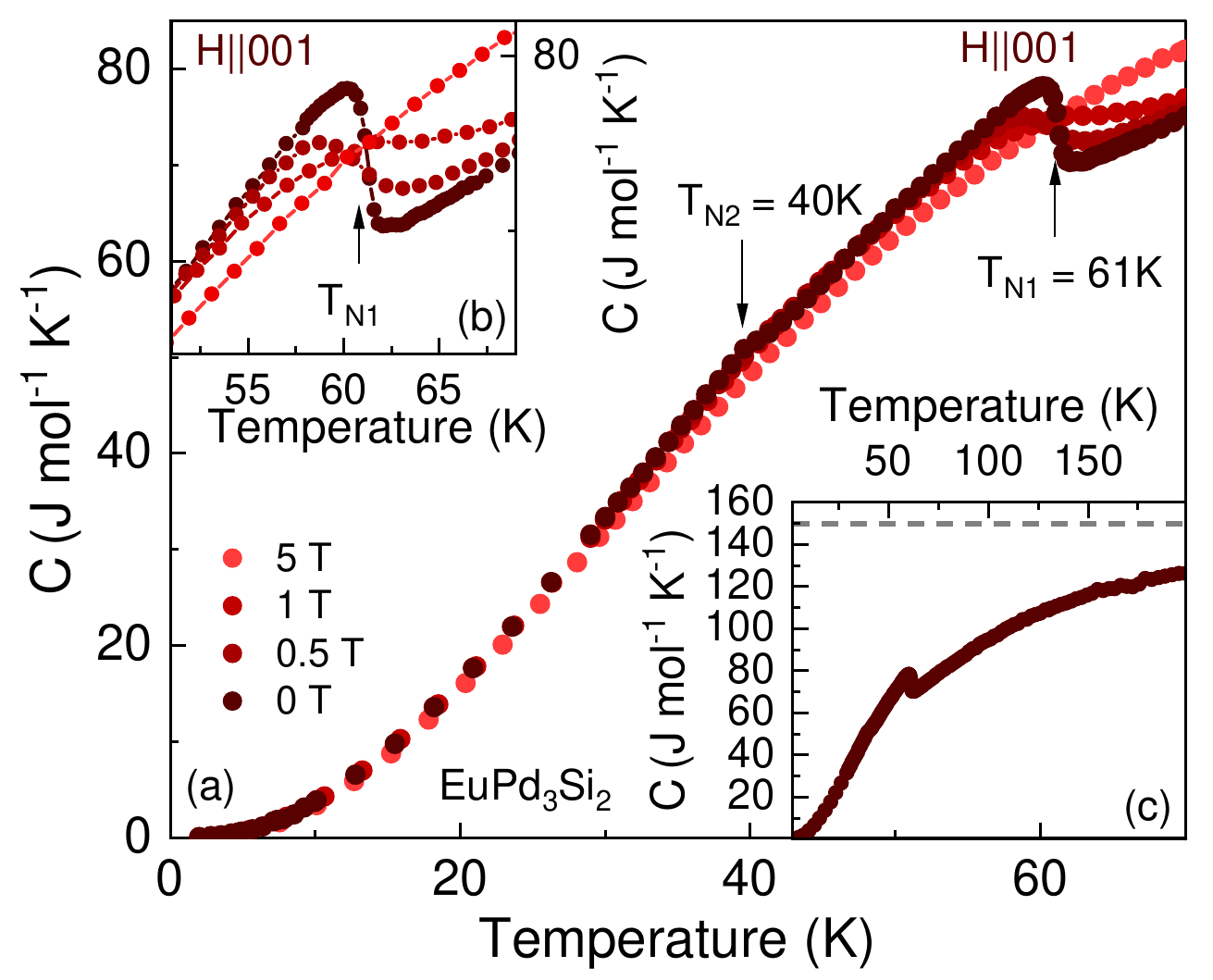}
    \caption{
    Heat capacity as function of temperature for different magnetic fields. The inset shows the measurement of the heat capacity as function of temperature up to 200 K. (b) The peak at $T_{\rm N1}$ shifts  to lower temperatures with higher magnetic fields. (c) At high temperatures, the heat capacity approaches the Dulong-Petit limit of 150 J/(mol K).}
    \label{fig:HC_EuPd3Si2}
\end{figure}
\noindent
Fig.~\ref{fig:HC_EuPd3Si2}(a) shows the heat capacity as a function of temperature for different magnetic fields applied along the $[001]$ direction. 
In zero magnetic field, the material shows two transitions. A sharp $\lambda$-type peak at $T_{\rm N1}=61~\rm K$ and a weak kink at $T_{\rm N2}=40~\rm K$.
As shown in Fig.~\ref{fig:HC_EuPd3Si2}(b), 
the peak shifts to lower temperatures with increasing magnetic field which is consistent with AFM order. In addition, the peak is smeared when applying higher magnetic fields.
Measurements of response to long heat pulses show that the transition is continuous and not of first order. The kink at 40\,K is again broadened in magnetic field and is no longer recognizable for $\mu_0H=5$\,T. \\
The Sommerfeld coefficient was determined from the data below 10\,K and results in a value of $\gamma$ = (48 $\pm$ 5)~mJ/molK. 
The data are in contrast to those presented for the compound EuPd$_3$Si$_2$ in Ref.~\cite{sharma2023crystal}, where two $\lambda$-type peaks have been observed at $T_{\rm C1}$ = 78\,K and $T_{\rm C2}$ = 5\,K, and from which it was concluded that a long-range order of Eu is already observed below $T_{\rm C1}$. 
\subsection{Resistivity}\noindent
\begin{figure}
    \centering
    \includegraphics[width=0.99\linewidth]{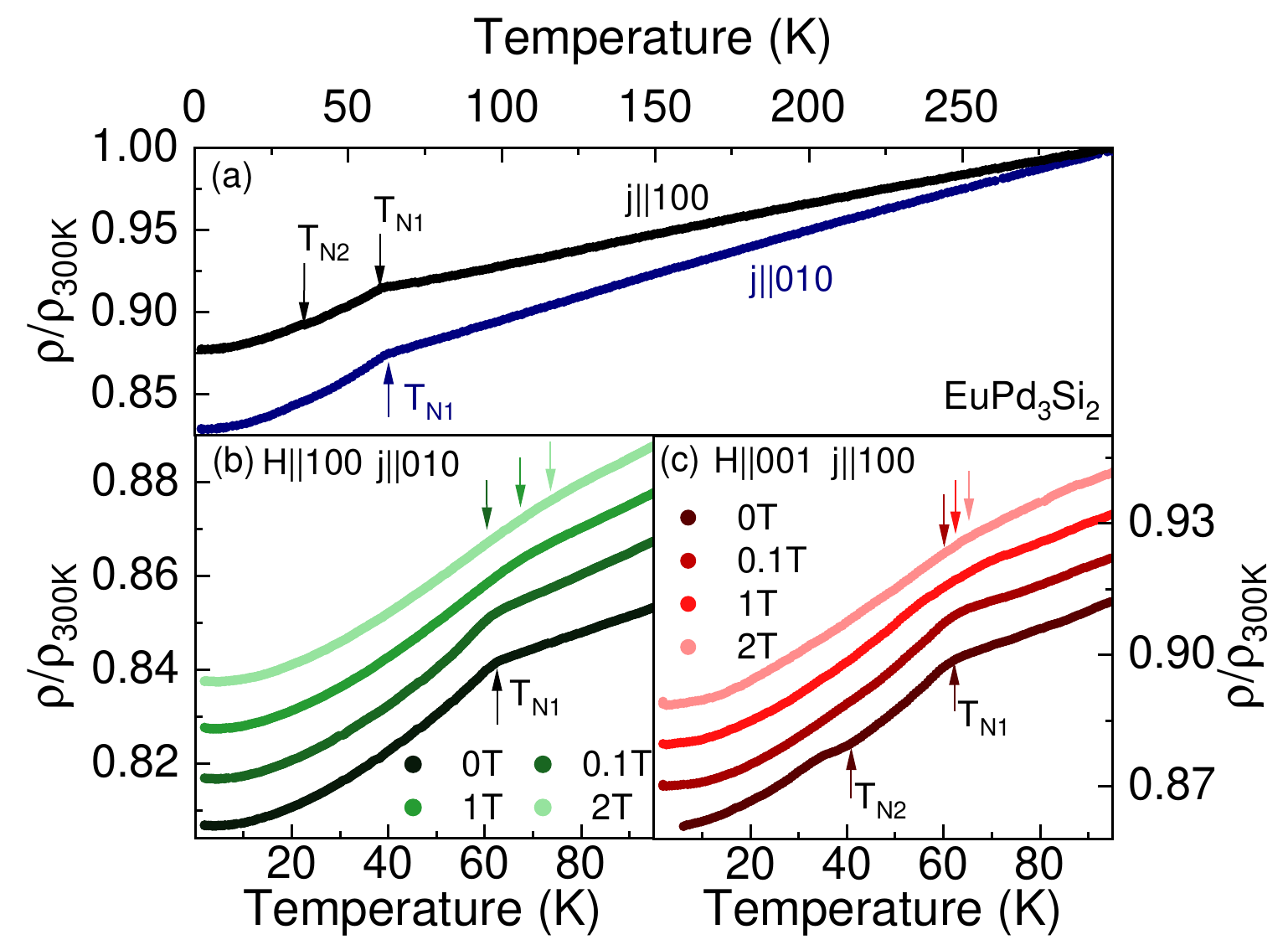}
    \caption{(a) Normalized electrical resistivity as function of temperature for the current applied along the $[100]$ and $[010]$ directions. Resistivity measured for different magnetic fields applied along the $[100]$ direction (b) and along the $[001]$ direction (c). The measurements performed under applied magnetic fields are shown with an offset of +0.01 for clarity.  }
    \label{fig:ACT}
\end{figure}\noindent
Resistivity as a function of temperature, $\rho(T)$, shown as normalized resistivity $\rho/\rho_{300\rm K}$ in Fig.~\ref{fig:ACT}(a), was measured between 300\,K and 2\,K with current applied along the $[010]$ (blue) and along the $[100]$ (black) directions.
With decreasing temperature, the resistivity decreases until a sudden drop appears at $T_{\rm N1}$ for both current directions. A further kink can be observed for $j\parallel[100]$ close to $T_{\rm N2}$.
Such a kink is reproducibly observed only during the measurement where the current was applied along the $[100]$ direction. Additional measurements on different samples are presented in Fig.~S2 in the supplement~\cite{supplementalinfo_EuPd3Si2_2025}. No anomaly was observed at $T_{\rm N2}$ when the current is applied along $[010]$. \\
With an applied magnetic field along $[100]$ (Fig.~\ref{fig:ACT}(b)) and along $[001]$ (Fig.~\ref{fig:ACT}(c)), the transition $T_{\rm N1}$ shifts only slightly toward lower temperatures, as indicated by the arrows. For fields of $\mu_{0}H \geq 1$\,T, the compound enters a field-polarized (FP) state for both field directions, which leads to a pronounced broadening of the transition and causes it to appear at higher temperatures. For a field applied along $[001]$ $T_{\rm N2}$ is no longer observable at a field of 0.1\,T.
The residual resistivity ratio for the measured samples of RR$_{\rm 2K}=\rho_{300\rm K}/\rho_{2\rm K} \approx$ 1.2 ($j\parallel[010]$) is small. A similar characteristic was observed in Ref.~\cite{sharma2023crystal} at different temperatures $T_{\rm C1}$ and $T_{\rm C2}$. The obtained RR$_{\rm 2K}$$\approx$1.5 is comparable to the value determined from the data taken from the samples of this study. The unusually low residual resistivity ratio may arise from the presence of magnetic domains or from Pd atoms occupying Si sites in the samples.

\subsection{Temperature dependence of the susceptibility}\noindent
Before delving into a detailed discussion of the anisotropic susceptibility, it is important to emphasize that long-range ferromagnetic order would typically be signified by certain key indicators. These would include a divergence in susceptibility at the transition, the emergence of spontaneous magnetization at low fields, as well as the presence of remanent magnetization or hysteresis in the magnetization $M(H)$ curve. However, in the current context, all of these phenomena can be clearly dismissed.
Susceptibility was measured as a function of temperature between 1.8\,K and 150\,K and is presented for a field of $\mu_0H=0.01\,\rm $T aligned along the three main symmetry directions in Fig.~\ref{fig:MvT_EuPd3Si2} and for higher fields in Fig.~S5 in the supplement \cite{supplementalinfo_EuPd3Si2_2025}. A strong anisotropy is observed for fields applied along the $[100]$, $[010]$ and $[001]$ directions, which is consistent with a magnetocrystalline anisotropy due to the orthorhombic structure of the material. The magnetic susceptibility shows a transition into the magnetically ordered phase at $T_{\rm N1}=61\,\rm K$ and a further transition at $T_{\rm N2}=40\,\rm K$. This is in contrast to the results presented in Ref.~\cite{sharma2023crystal}, where a transition at $T_{\rm C1}=78\,\rm K$ and a spin reorientation at $T_{\rm C2}=5~\rm K$ were reported for the same material (gray symbols in Fig.~\ref{fig:MvT_EuPd3Si2}). 
The authors of Ref.~\cite{sharma2023crystal} deduced FM order from neutron powder diffraction measurements but also mention that their magnetization data are in accordance with a canted AFM-type of order.\\ 
For the samples studied here, AFM order was derived from the characteristic susceptibility curves \cite{Coey2010}. When the magnetic field is applied parallel to the direction of the magnetic moments in an AFM material, it tends to align the moments more easily. In this configuration, the susceptibility is generally higher because the external field can effectively couple with the aligned moments. According to mean field theory, when the external field is applied parallel to the magnetic moments, the field reduces the energy barrier for flipping spins, and the system can respond more readily, enhancing the susceptibility often denoted as $\chi_{\parallel}$.   When the magnetic field is applied perpendicular to the direction of the magnetic moments, the interactions between the spins remain more robust. The antiferromagnetic coupling mechanism tends to resist the alignment due to the field. In this case, the susceptibility is usually lower compared to when the field is parallel to the moments. The perpendicular field does not effectively couple with the spins due to their preferred antiparallel arrangement and thus does not lead to a strong response. The system may still exhibit some weak response as the spins can slightly cant out of their equilibrium positions, but it will be significantly less than the parallel case.  The mean field theory predicts that the susceptibility is significantly reduced in the perpendicular alignment due to the stability of the AFM ordering, and therefore the spins are not easily aligned by the magnetic field, leading to lower values of susceptibility often denoted as $\chi_{\perp}$.

Our susceptibility measurements show that between 61\,K and 40\,K, the susceptibility is constant for $H\parallel [010]$ and $H\parallel [001]$ which is a hallmark of collinear antiferromagnetic order within mean-field theory indicating that the field is aligned perpendicular to the moments in these cases. The susceptibility can be denoted as $\chi_{\perp}$ in this $T$ range. In case of $H\parallel [100]$, a $\chi_{\parallel}$ behavior was found with a positive slope of $\chi$ between 61\,K and 52\,K while between 52\,K and 40\,K the slope is negative. This particular temperature dependence might be related to the redistribution of different AFM domains for $H\parallel [100]$ between $T_{\rm N1}$ and $T_{\rm N2}$. Upon heating from 2\,K to $T_{\rm N2}=40\,\rm K$, a change in $\chi(T)$ occurs that is most pronounced for $H\parallel [010]$ while $\chi(T)$ is high and almost  constant for $H\parallel[100]$. This behavior below $T_{\rm N2}$ is assigned to a canting of the magnetic moments in the material towards the $[010]-[001]$ plane away from the $[100]$ direction. Below $T_{\rm N2}$ for all applied field directions, a hysteresis between the heating and cooling curves occurs, which could be caused by a redistribution of magnetic domains in the material. Based on the temperature-dependent data, the following picture emerges: Below $T_{\rm N1}$, the moments align to a configuration parallel to $[100]$ along the Eu chains. Between 2\,K and $T_{\rm N2}$, the moments tilt away from the $[100]$ direction towards the $[010]-[001]$ plane.\\
Furthermore, we studied the paramagnetic phase to get insights about dominant magnetic interactions in the material. The inset in Fig.~\ref{fig:MvT_EuPd3Si2} shows the inverse susceptibility as a function of temperature for a field of $\mu_0H=1\,\rm T$ aligned along the $[010]$ direction. The orange line indicates the Curie-Weiss fit in the temperature range between 300\,K and 100\,K.  Here, the compound shows an effective paramagnetic moment of $\mu_{\rm eff} = (8.3\pm 0.4)\,\mu_{\rm B}$ and a Weiss temperature of $\Theta_W$ = (60.5 $\pm$ 0.5)\,K. 
The effective moment $\mu_{\rm eff}$ is slightly larger than the calculated moment of Eu$^{2+}$ (7.94\,$\mu_{\rm B}$), but both values are in good agreement with those reported in Ref.~\cite{sharma2023crystal}. A slightly enhanced effective magnetic moment can be caused by additional contributions originating from outer 5d or 6s Eu electrons \cite{Koebler1975}. At first glance, a positive Weiss temperature as determined here  could suggest dominant ferromagnetic interactions, but the Weiss temperature reflects the sum of all interactions in a material. 
The results of the susceptibility measurements point to an arrangement of the magnetic moments in EuPd$_3$Si$_2$ in FM chains along the [100] direction, which are AFM coupled. Furthermore, $T_{\rm N1}=61\,\rm K$ of our EuPd$_3$Si$_2$ samples is almost equal in absolute value to $\Theta_W$. The observation of a FM Weiss temperature very similar in absolute value to the AFM ordering temperature was made earlier for instance in case of EuRh$_2$Si$_2$ \cite{seiro2013complex}. Here, it was argued that this points to a strong coupling within each FM sublattice and a much weaker intersublattice coupling.
\\
The temperature dependent data recorded in higher fields, see Fig.~S5 in Ref.~\cite{supplementalinfo_EuPd3Si2_2025}, show that the transition at $T_{\rm N1}$ shifts to lower temperatures with an increasing field applied along the main symmetry directions, which is consistent with AFM order. In addition, the field that is needed to reach the field-polarized state is strongly anisotropic. While the transition 
broadens already at a low field of $\approx 0.3$\,T, for the field along $[100]$, Fig.~S5(a), it is clearly visible up to a field of 2.5\,T applied along $[001]$, Fig.~S5(c). The transition at $T_{\rm N2}$, which is assigned to a canting of the magnetic moments in the structure toward the $[010]-[001]$ plane, is clearly visible with a field applied along the $[010]$ and $[001]$ directions. Similar to the transition at $T_{\rm N1}$, it decreases in temperature with an increasing field and is smeared out in a low field of $0.05\,\rm T$ for $H\parallel[100]$ and a higher field of $\approx 1\,\rm T$ for $H\parallel[010],[001]$, respectively. 

\begin{figure}[ht!]
    \centering
    \includegraphics[width=0.47\textwidth]{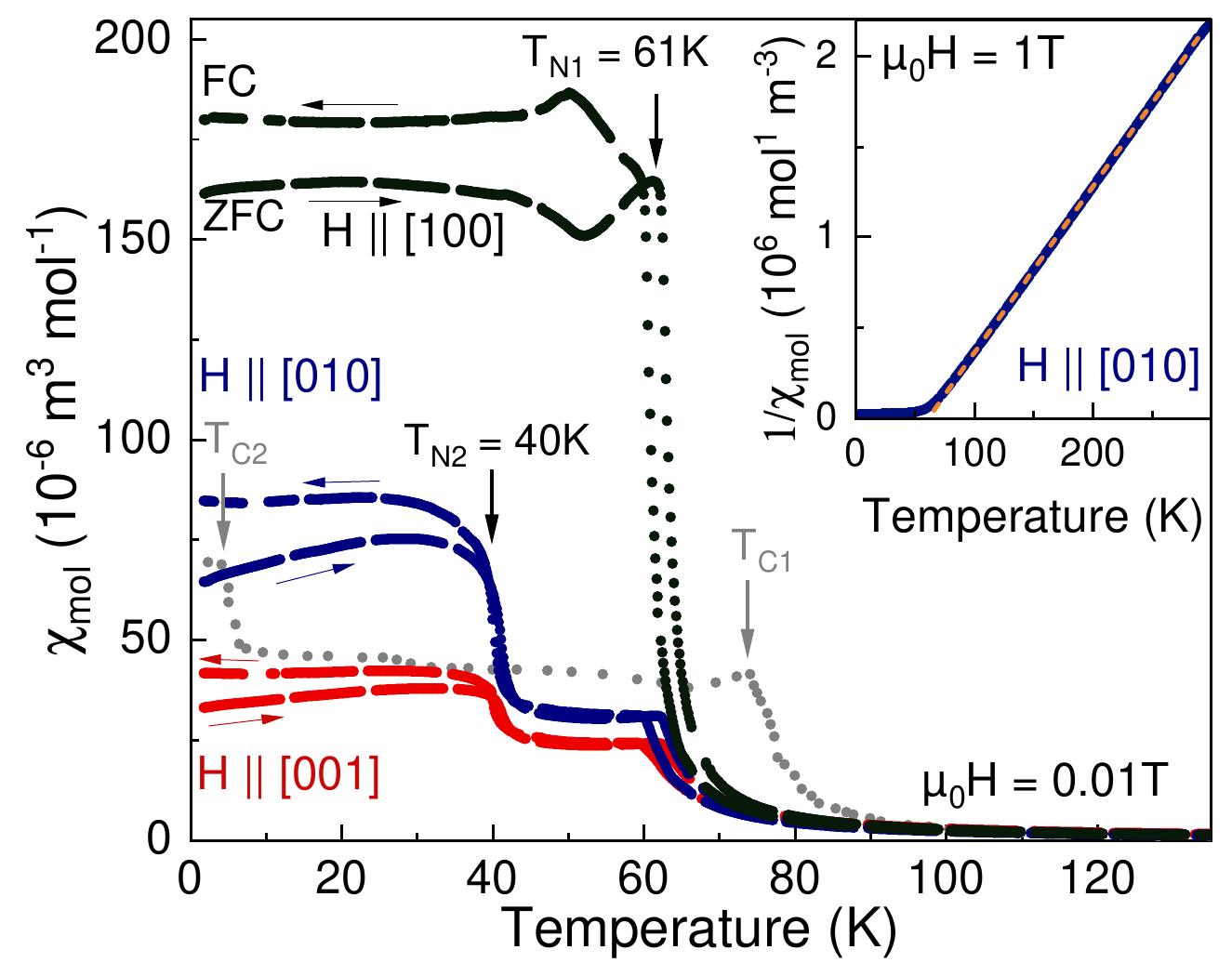}
    \caption{Magnetic susceptibility as function of temperature for different directions of an applied field of 0.01\,T. Two transitions are observed. The data in gray are taken from \cite{sharma2023crystal}. The inset shows the temperature dependent inverse susceptibility measured at 1T with the field applied alog the $[010]$ direction. The orange line indicates a Curie-Weiss fit to the data in the temperature range 100\,K $\leq\, T \leq$ 300\,K. }
    \label{fig:MvT_EuPd3Si2}
\end{figure}

\subsection{Field dependence of the magnetization}
\begin{figure}[ht!]
    \centering
    \includegraphics[width=0.47\textwidth]{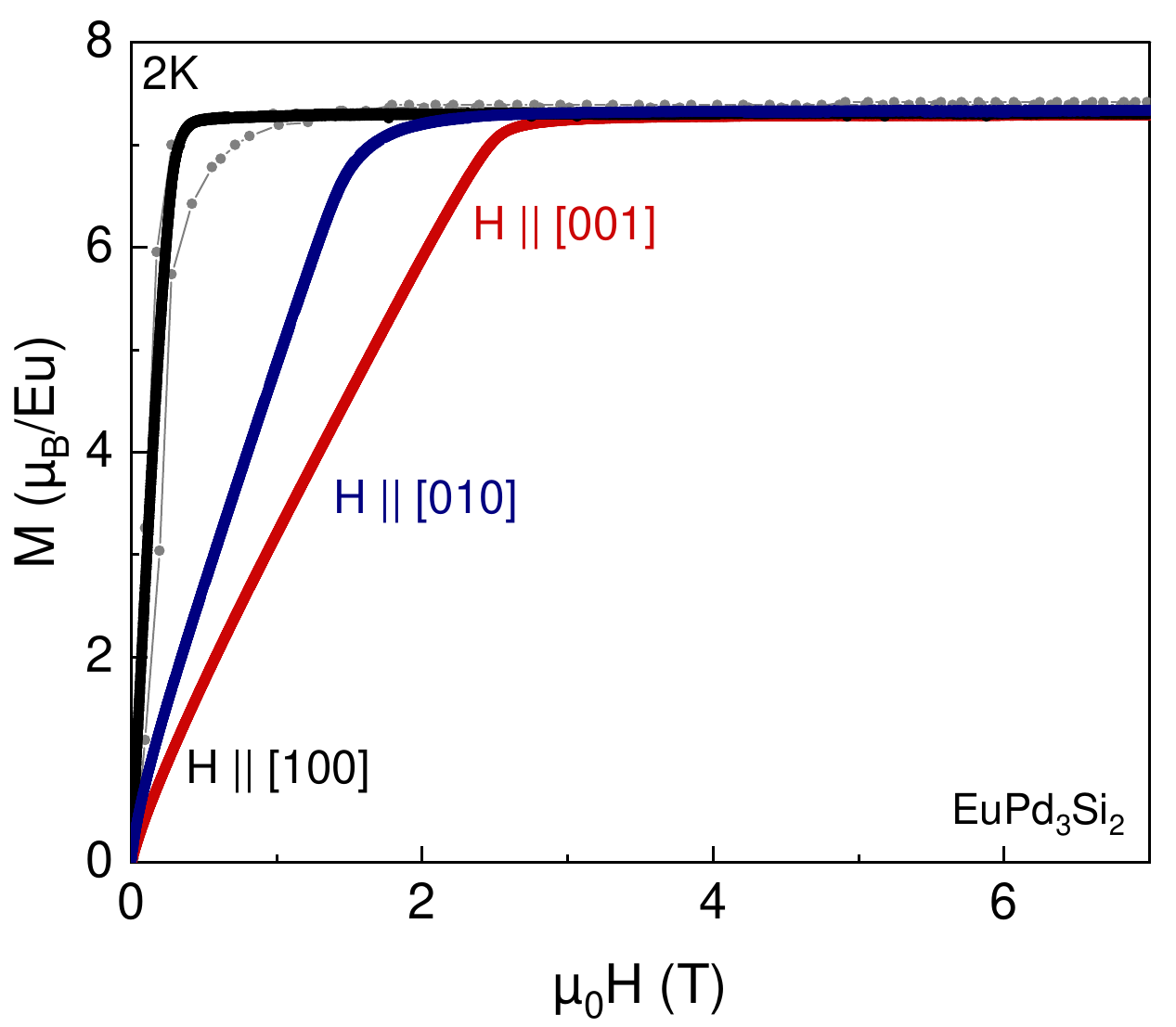}
    \caption{ Magnetic moment per Eu as function of the magnetic field for field applied along the three main symmetry directions at 2\,K. 
    The data in gray are taken from \cite{sharma2023crystal}.}
    \label{fig:MvH_EuPd3Si2}
\end{figure}

\begin{figure}[ht!]
    \centering
     \includegraphics[width=0.47\textwidth]{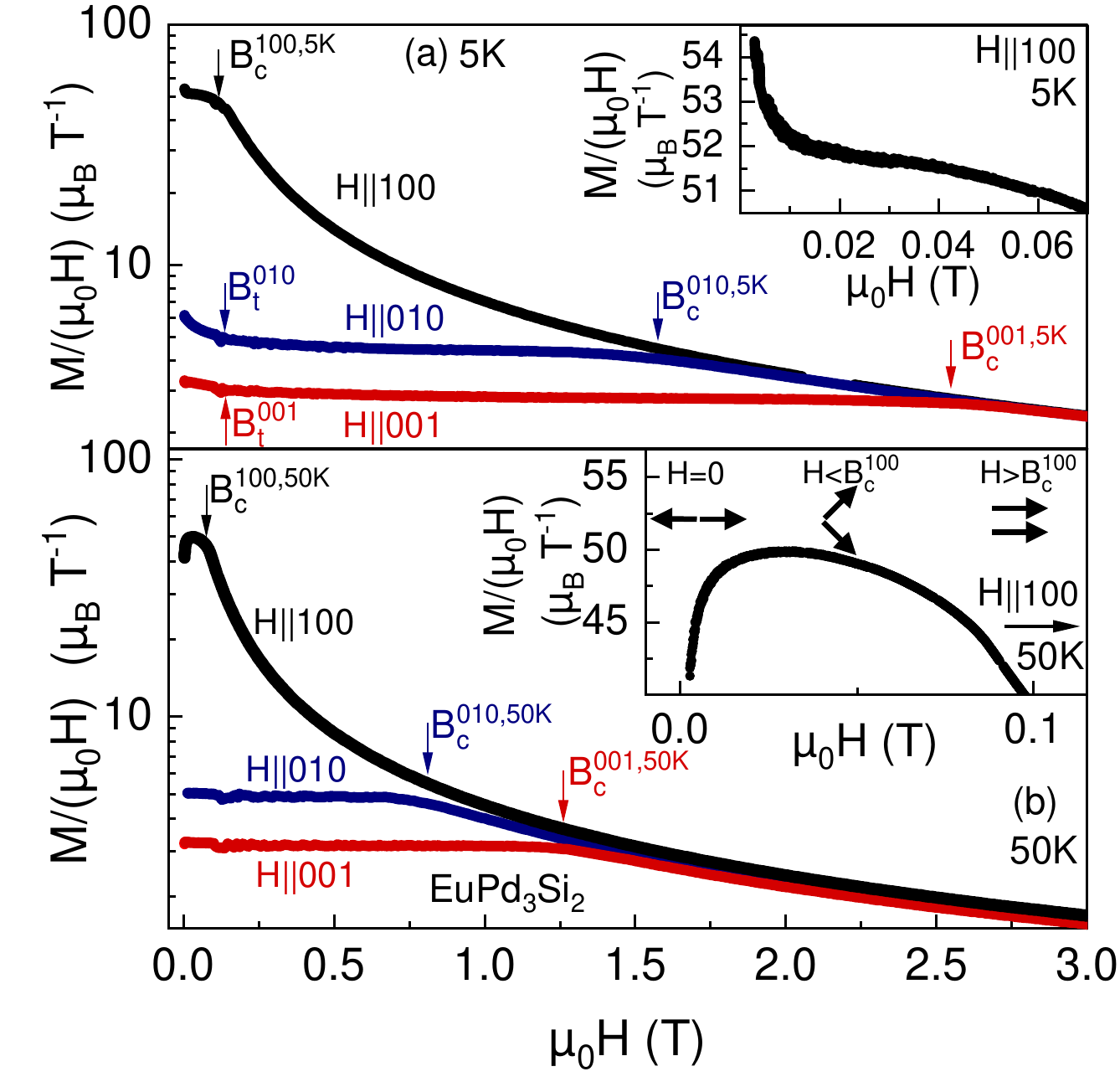}
     \caption{ Comparison of the behavior of $M/H$ versus $\mu_0H$ at low fields along the three main directions at (a) $T = 5~\rm K$ and at (b) 50\,K. }
    \label{fig:MdurchH}
\end{figure}

\noindent
In accordance with the susceptibility data, the magnetic moment per Eu as a function of the magnetic field at 2\,K in Fig.~\ref{fig:MvH_EuPd3Si2} shows a strong anisotropy for fields applied along the three main directions with a saturated moment of$M^{\rm sat}$ = 7.34~$\mu_{\rm B}$ per Eu which is slightly higher than the expected value for Eu$^{2+}$ of $M^{\rm sat}$ = $g_JJ=7~\mu_{\rm B}$. 
This by $\Delta M^{\rm sat}\approx 0.3\,\mu_{\rm B}$ at 2 K enhanced moment is consistent to what was observed in \cite{sharma2023crystal} where band structure calculations hint to the polarization of the valence band being responsible for the enhanced value of the saturated moment. The saturation fields were determined by the derivative of the magnetization, $\mathrm{d}M/\mathrm{d}H$, shown in the supplemental material \cite{supplementalinfo_EuPd3Si2_2025} in the inset of Fig.~S4(c). The moment saturates in the lowest critical field for $H\parallel [100]$ of $B_c^{100,2\rm K}=0.3~\rm T$, for $H\parallel [010]$ at $B_c^{010,2\rm K}=1.6~\rm T$ and finally for $H\parallel [001]$ at $B_c^{001,2\rm K}=2.5~\rm T$. For comparison, the $M(H)$ data taken from \cite{sharma2023crystal} are shown in gray in the main figure, which show good agreement with this study. There, no direction of the magnetic field was given. However, from the comparison with our data in Fig.~\ref{fig:MvH_EuPd3Si2}, we can infer that the field was applied along the [100] direction.
At $50\,\rm K$ we as well find a reduced $M^{\rm sat}$ = 5.2\,$\mu_{\rm B}$. As shown by band structure calculations in \cite{sharma2023crystal}, the polarization of the valence band might cause an additional contribution to the saturation moment at low temperatures. One can also reasonably expect that, also at high temperatures, this polarization might influence the saturation value. The reduction in $M^{\rm sat}$ at high temperature might then occur if the net polarization of the valence band and that of the localized Eu 4$f$ moments are not aligned.\\ \noindent
In contrast to the data shown in \cite{sharma2023crystal}, no hysteresis was observed at low magnetic fields. The magnetic moment per Eu atom as a function of the magnetic field in the low-field region is shown in Fig.~S3 in the supplemental material \cite{supplementalinfo_EuPd3Si2_2025}. Neither at 50\,K nor at 2\,K a hysteresis is detected in $M(H)$. At higher temperatures, the critical fields shift to lower values. Data recorded at different temperatures with fields along the different crystallographic axes are attached as Fig.~S4 in the supplement \cite{supplementalinfo_EuPd3Si2_2025}.\\
From the field dependence of the magnetization, further information on the moment alignment in an antiferromagnet can be obtained as it was demonstrated previously \cite{Kliemt2017}. 
For better comparison of the data below and above $T_{\rm N2}$, $M/H$ as a function of the magnetic field is shown in Fig.~\ref{fig:MdurchH}.
In this representation, changes in slope at low fields indicate moment reorientation, whereas the high-field regime corresponds to the approach of the field-polarized state along different directions.\\
At low fields and 5\,K, Fig.~\ref{fig:MdurchH}(a), data measured with the field applied along the $[100]$ direction show a small change of slope below $\mu_0H=0.01\,\rm T$ followed by a plateau below $B_c^{100,5\rm K}$ at high absolute values.
The slope change can be attributed to reorientations of magnetic moments in the field into a "moments perpendicular to the field" state. The constant $M/H$ versus $H$ plateau indicates that the moments that are aligned perpendicular to the field tilt toward the field. 
In contrast, $M/H$ is low, and exhibits a small slope change below $B_t^{010}$ and $B_t^{001}$. 
When the field-polarized state is reached, $M/H$ decreases. At 5\,K, the material becomes field polarized along $[100]$ in the lowest critical field $B_c^{100,5\rm K}$, the FP state for the field along $[010]$ and along $[001]$ is reached at higher fields $B_c^{010,5\rm K}$ and $B_c^{001,5\rm K}$. This particular field dependence suggests that in zero field at $T=5\,\rm K$ the magnetic moments are aligned in the structure along the $[100]$ / $[\bar{1}00]$ direction but slightly tilted toward $[010]$ and $[001]$.    \\
At 50\,K, Fig.~\ref{fig:MdurchH}(b), the field dependence changes slightly. For $H\parallel [100]$, a change in slope is visible below $0.01\,\rm T$. Also, here the low-field absolute values are high, indicating that a moment reorientation occurs exclusively along the $[100]$ direction. Here, the field-polarized state is reached at a lower field and the plateau of "moments tilting toward the field" is much smaller than at 5\,K. This might mean that at 50\,K the moments almost immediately switch to the field-polarized state in extremely low magnetic fields meaning that the material is easily polarizable.\\
Considering this overall field dependence at 50\,K the moments are mainly aligned along the Eu chains that lie along the $[100]$ direction in the crystal structure (see the schematic drawing in the inset of Fig.~\ref{fig:MdurchH}(b)). We do not find a clear change of slope in $M/H$ at low fields but nevertheless we cannot exclude the presence of a minor canting away from $[100]$ below $T_{\rm N1}$ as proposed in \cite{sharma2023crystal}. At 5\,K, this orientation appears to have changed, leading to a slight canting of the moments away from $[100]$ toward the $[010]$ and $[001]$ directions.


\subsection{Phase diagram}
\begin{figure}
    \centering
    \includegraphics[width=1\linewidth]{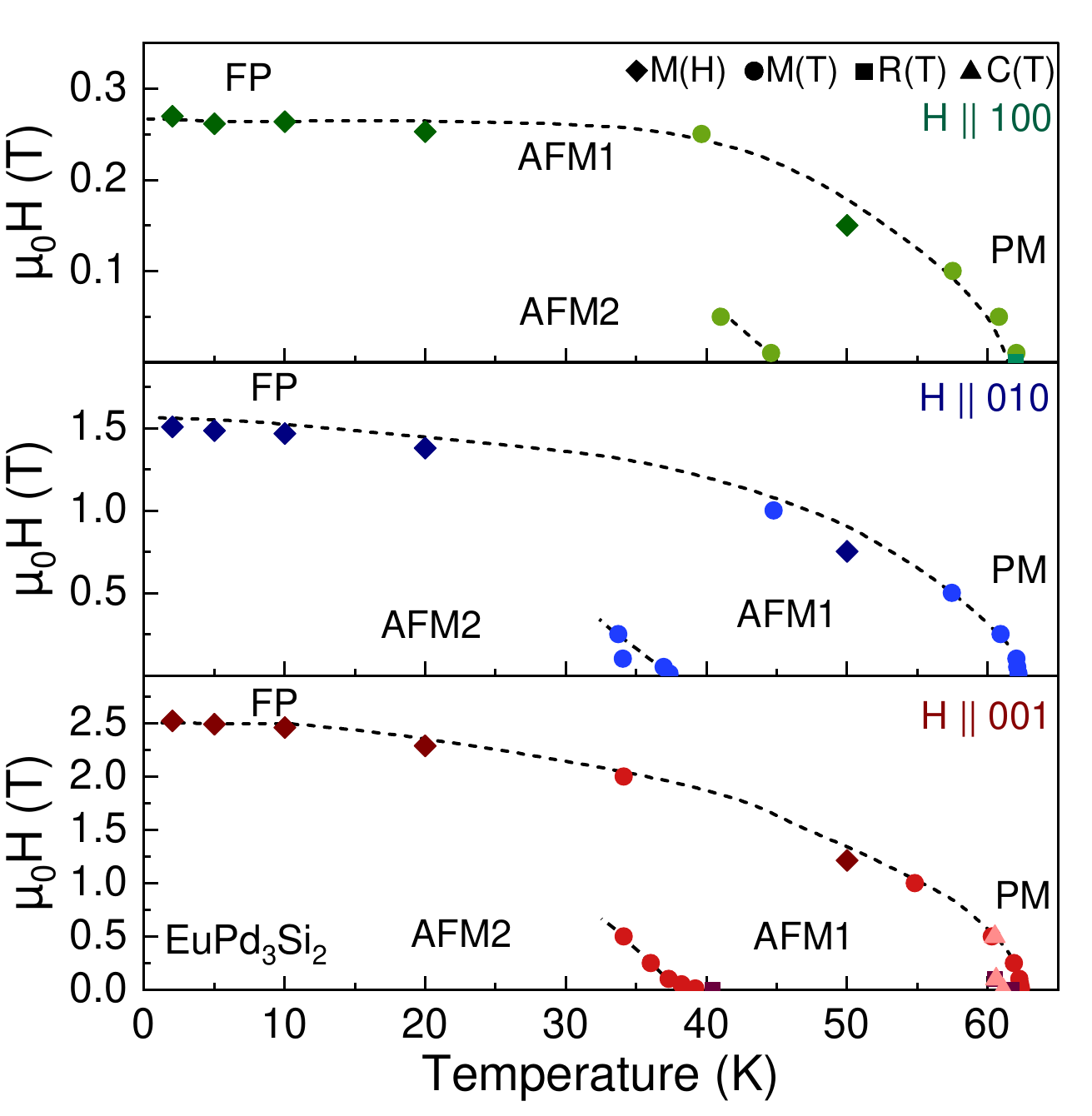}
    \caption{ Field-temperature phase diagram for the three main symmetry directions. The data were extracted from temperature-dependent susceptibility (circles), resistivity (squares), heat capacity (triangles) and field-dependent magnetic moment (diamonds). The dashed lines are guides to the eyes. }
    \label{fig:Pd_EuPd3Si2}
\end{figure}

\noindent
From temperature-dependent susceptibility (circles), resistivity (squares), heat capacity (triangles) and field-dependent magnetic moment (diamonds) phase diagrams as shown in Fig.~\ref{fig:Pd_EuPd3Si2} were constructed. Data were extracted from the plots in Fig.~S4 and Fig.~S5 in Ref.~\cite{supplementalinfo_EuPd3Si2_2025}.
$B_c^{100}$ was obtained from the data of Fig.~S4 plotted as $dM/dH$ as function of magnetic field as shown the inset of Fig.~S4(c) in supplemental material \cite{supplementalinfo_EuPd3Si2_2025}.
$T_{\rm N1}$ and $T_{\rm N2}$ were determined from the intersections of two linear regressions (Fig.~S5 in Ref.~\cite{supplementalinfo_EuPd3Si2_2025}). In the vicinity of $T_{\rm N1}$, the susceptibility as function of temperature exhibits a thermal hysteresis, which is why the zero-field-cooled (ZFC) curves were used to determine the transition temperatures. \\
With decreasing temperature, the compound undergoes a transition from the paramagnetic state to the magnetic order (AFM1) at $T_{\rm N1}$ in which the moments are probably mainly aligned along the $[100]$. A further transition which is assigned to a spin reorientation where the moments are probably further canted toward the $[010]-[001]$ plane (AFM2 phase) appears at $T_{\rm N2}$.
Different field strengths are required along the three main directions to achieve the FP state. While the smallest fields are needed along the $[100]$ direction to align the magnetic moments in the field, much larger fields are required along the $[001]$ direction. 

\section{Discussion}\noindent

In this study, EuPd$_3$Si$_2$ single crystals were investigated and their properties were compared to those recently reported for crystals of the same material in Ref.~\cite{sharma2023crystal}.
A difference between the two studies lies in the crystal growth procedure, which differs slightly with respect to initial melt stoichiometry, crucible setup and maximum temperature. While in \cite{sharma2023crystal} EuPd$_3$Si$_2$ was grown as a side phase from the initial melt stoichiometry of  Eu:Pd:Si=1:2:2, an optimized stoichiometry of Eu:Pd:Si=1.45:2.5:1.5 was used in the present study, from which phase pure EuPd$_3$Si$_2$ as the main phase was obtained. The decisive difference in the construction of the crucible setup was the use of a graphite inner crucible instead of the tantalum crucible used in \cite{sharma2023crystal}. To minimize the contamination of the melt by the crucible material, a lower maximum temperature was chosen than that reported in \cite{sharma2023crystal}. 
As described in Ref~\cite{sharma2023crystal}, the Ta crucible was partially dissolved by the melt, resulting in the formation of TaSi$_2$ as a side phase.
This loss of Si affected not only the initial melt stoichiometry but also the use of Ta could have led to unintended substitution of the crystals used there. This possible 
substitution can affect the nature of the magnetic transition. \\
In addition, SC-XRD shows a $\approx$ 7 \% excess of Pd substitution on the Si site. A change in the Pd–Si ratio can slightly modify the local environment and the crystal field of the Eu atoms, thereby influencing the material’s physical properties, as observed in EuPd$_2$Si$_2$ \cite{Kliemt2022a}.  
A similar case of growth condition-dependent physical properties was observed for the compound EuCd$_2$As$_2$, in which an antiferromagnetic (crystal growth from Sn flux in Al$_2$O$_3$ crucibles \cite{Jo2020}) or ferromagnetic ground state (crystal growth from salt flux in SiO$_2$ ampoules \cite{Jo2020}) was observed. 
A possible explanation for the different magnetic orders can be attributed to variations in interatomic distances. The Eu–Eu distance along the chains, determined from SC-XRD d$_1$=3.6258(1)\,\AA, is larger than the distance d$_{1\ast}$=3.5962\,\AA\, extracted from the data reported in Ref.~\cite{sharma2023crystal}. The Eu bond lengths within the $b$–$c$ plane, d$_{\rm 2}$~=~5.7393(5)\,\AA, d$_{\rm 3}$~=~6.0062(3)\,\AA, and d$_{\rm 4}$ = 5.5800(3)\,{\AA}, also differ from those reported in Ref.~\cite{sharma2023crystal}, where the corresponding values are d$_{\rm 2*}$ = 5.7506 {\AA},  d$_{\rm 3*}$ = 6.0072 {\AA},  d$_{\rm 4*}$ = 5.6134 {\AA}.  These differences, likely induced by some form of substitution, can alter interactions between Eu atoms, as has been observed in other compounds \cite{nikitin2002influence,paschen2001structural}. 
\\
The magnetic anisotropy observed in compounds containing rare-earth ions mainly results from spin-orbit coupling and crystal field effects. In the case of EuPd$_3$Si$_2$, the spin-orbit coupling has little impact on the $4f$ shell of Eu$^{2+}$ due to L = 0. The low anisotropy seen in the exchange, attributed to the difference $ \Theta^a_W - \Theta^c_W \approx 4\,\rm K $, is likely due to spin-orbit coupling influencing the Pd states near the Fermi level. This interaction promotes spin polarization among the neighboring Eu moments via the RKKY exchange mechanism. In solid-state systems, the symmetry of the lattice around the rare-earth ion can result in different energy levels for various spin orientations. Despite  L = 0, the arrangement of adjacent atoms (for instance, Pd) generates a crystal field that can affect the energy levels of spin states. Therefore, the easy axis may differ based on the local environment and symmetry, leading to a preferred alignment of spins in specific directions, thus producing magnetic anisotropy.
The signatures of putative antiferromagnetic order observed in the present study below the transition temperatures $T_{\rm N1}$=40\,K and $T_{\rm N2}$=61\,K clearly differ from the ferromagnetism reported in \cite{sharma2023crystal}, which occurs at the transition temperatures $T_{\rm C1}$=78\,K and $T_{\rm C2}$=5\,K. In the present case, the RKKY interaction $J_{\rm RKKY}\propto \cos(2k_FR)/R^3$ determines the magnetic properties. Due to the change of the lattice parameters and bond distances compared to Ref.~\cite{sharma2023crystal}, it can be assumed that in the present case the coupling constant $J_{\rm RKKY}$ is close to a zero point of the cosine, where the type of magnetic order can already vary due to such changes.
Alternatively, another scenario is possible, when assuming that there are several $J_{\rm RKKY}$, which is rather likely in this crystal structure. Then, the different ground states would result from competing FM and AFM interactions: In the present study, $J_{\rm RKKY}^{\rm AFM}$ is found to be dominant, whereas in Ref.~\cite{sharma2023crystal} the prevailing $J_{\rm RKKY}$ might favor FM alignment.
This would also explain the different ordering temperatures. 
A comparative investigation of both samples could provide valuable information on these differences.\\
\section{Summary}\noindent
Single crystals of the compound EuPd$_3$Si$_2$ were grown from a Pd-rich melt. 
The EDX analysis shows that the stoichiometry of the crystals matches that of the ideal 1:3:2 composition within the expected systematic and statistical errors of the method.\\
The result of the structural characterization showed that the EuPd$_3$Si$_2$ phase with  significantly different lattice parameters compared to those reported in Ref.~\cite{sharma2023crystal} was obtained. In particular, the $a$ lattice parameter of the crystals is $\approx$ 0.8\% larger and the $b$ lattice parameter $\approx$ 0.6 \% smaller than that reported in \cite{sharma2023crystal} but the particular reason for this difference is unclear. It could be caused by a difference in the Pd-Si ratio (different site occupancy deviations) in both types of samples or by a possible unintended doping with Ta in the samples investigated in \cite{sharma2023crystal}. The larger (smaller) lattice parameter $a$ ($c$) results in increased (reduced) Eu–Eu distances along (perpendicular to) the Eu chains, which may influence the RKKY interaction(s) between the ions. \\
Although we did not perform an explicit determination of the magnetic structure using neutron diffraction, we found key indications for AFM order in EuPd$_3$Si$_2$. The heat capacity and the resistivity show a slight shift of the magnetic transition to lower temperatures with increasing field. The strongest arguments for antiferromagnetism in this material is provided by the anisotropic susceptibility as the temperature-independent $\chi_{\perp}$ below $T_{\rm N1}$ is a hallmark of collinear antiferromagnetic order within mean-field theory. Such a behaviour would not be observed in case of FM order.
The temperature-dependent susceptibility gives evidence that antiferromagnetic ordering occurs at $T_{\rm N1}$=61\,K, followed by a spin reorientation transition at $T_{\rm N2}$=40\,K. These results differ from previously published ferromagnetic ordering in EuPd$_3$Si$_2$ at a higher temperature \cite{sharma2023crystal}. \\
Further evidence for an absence of FM order comes from the detailed study of the field dependence of the magnetization where no indications for a hysteresis were observed. With the field applied, the compound EuPd$_3$Si$_2$ shows anisotropy along the three main symmetry directions, which is summarized in phase diagrams in Fig.~\ref{fig:Pd_EuPd3Si2}. For small fields and low temperatures, the moments seem to cant away from the $[100]$ direction towards the $[010]$ and $[001]$, which is labeled as the AFM2 state in the phase diagram. With increasing temperature, the moments re-align along the Eu chains (AFM1). As the field and the temperature increase, the compound gradually enters the field-polarized state, first along the [100] direction, followed by [010], and finally along [001].\\
Besides EuPd$_2$Si$_2$ \cite{Kliemt2022a} and EuCd$_2$As$_2$ \cite{Jo2020}, EuPd$_3$Si$_2$ represents another Eu-based material where the properties of the magnetic ground state are sensitively dependent on the crystal growth conditions. Since the type of magnetic order is highly dependent on small changes in the lattice parameters, the material offers the possibility of switching the magnetic ground state through application of strain in future experiments.

\section*{Data Availability}
The data sets are available through https://doi.org/10.25716/gude.0k6n-7bfc in the open data repository of the Goethe University Frankfurt (GUDe). Structural data are available from the Karlsruhe Institute of Technology repository KITOpen via https://doi.org/ [zz].
\\
\begin{acknowledgments}

We are indepted to Siegmar Roth and Andre Beck at the Institute for Quantum Materials and Technologies, Karlsruhe Institute of Technology and to Tim F\"orster at the Goethe University for their great technical assistance. We acknowledge funding by the Deutsche Forschungsgemeinschaft (DFG, German Research Foundation) via the TRR 288 (422213477, projects A03 and B03).
\end{acknowledgments}
\bibliography{Lit}

\pagestyle{empty}

\begin{figure*}
\centering
\includegraphics[page=1,width=1.1\textwidth]{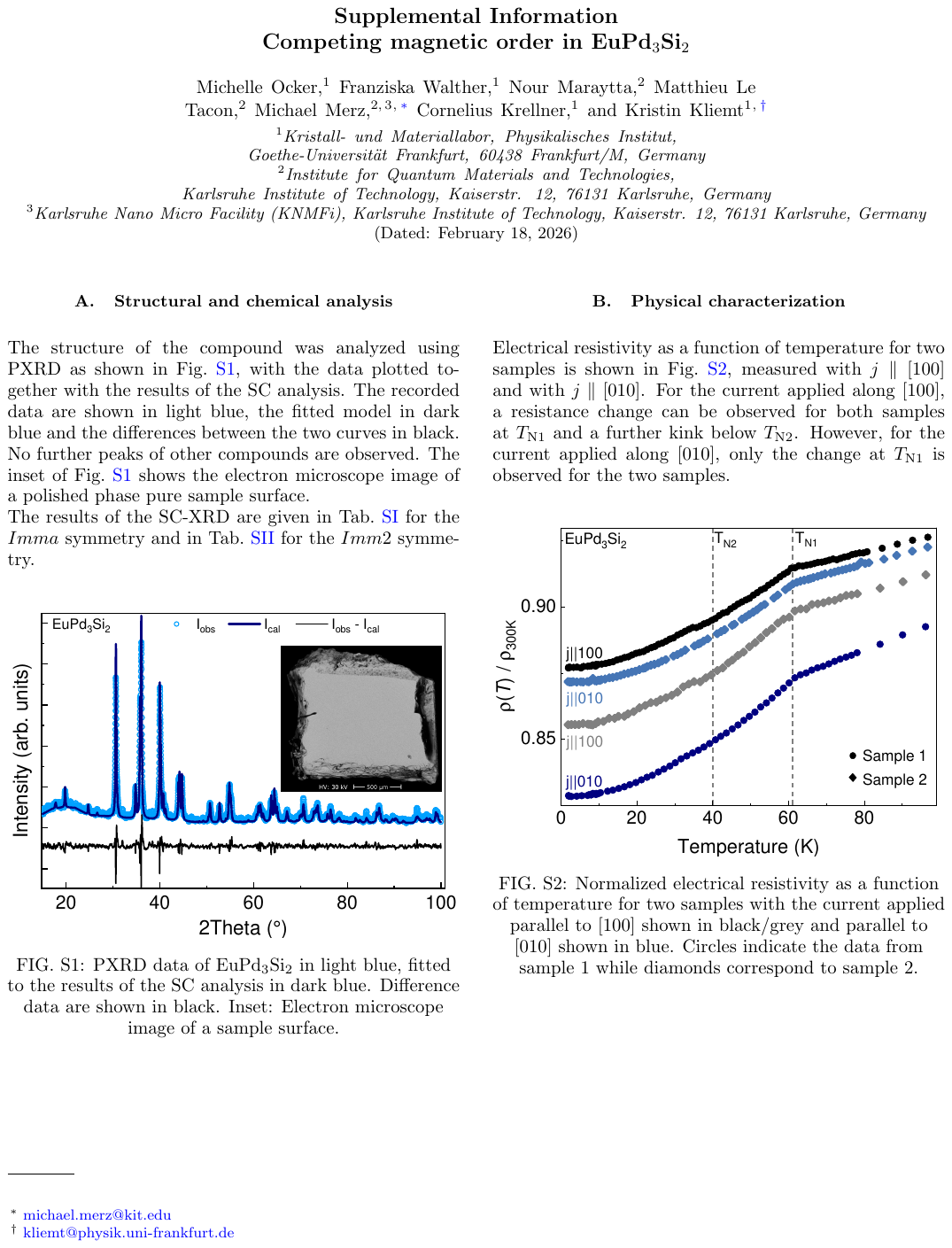}
\end{figure*}

\begin{figure*}
\centering
\includegraphics[page=2,width=1.1\textwidth]{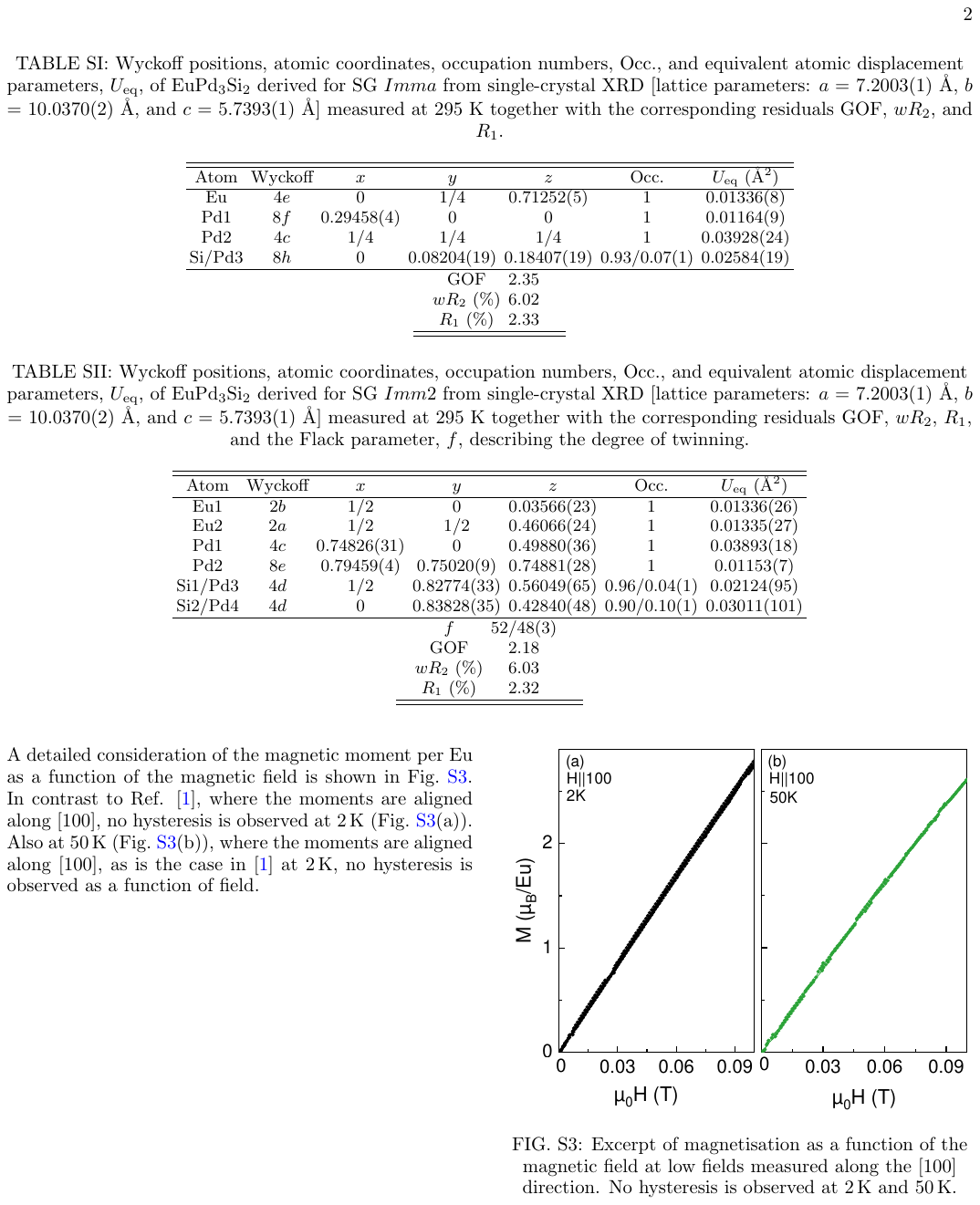}
\end{figure*}

\clearpage

\begin{figure*}
\centering
\includegraphics[page=3,width=1.1\textwidth]{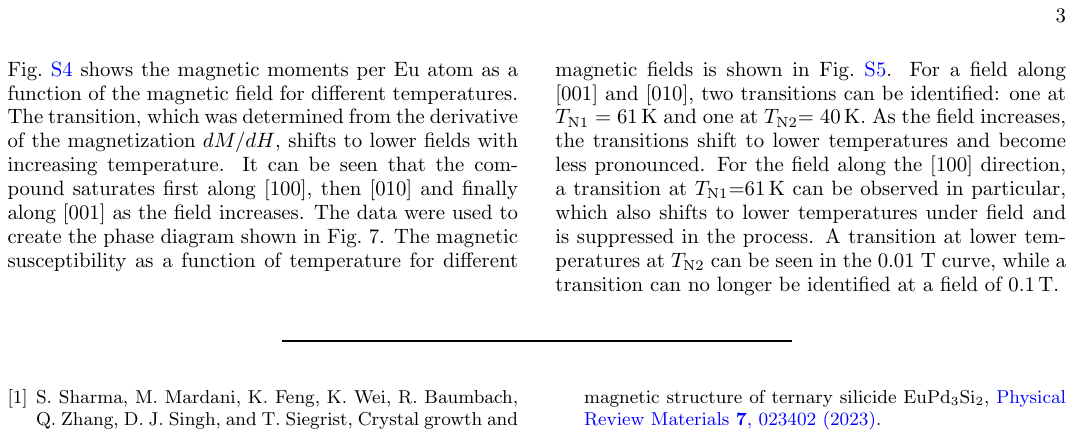}
\end{figure*}

\begin{figure*}
\centering
\includegraphics[page=4,width=1.1\textwidth]{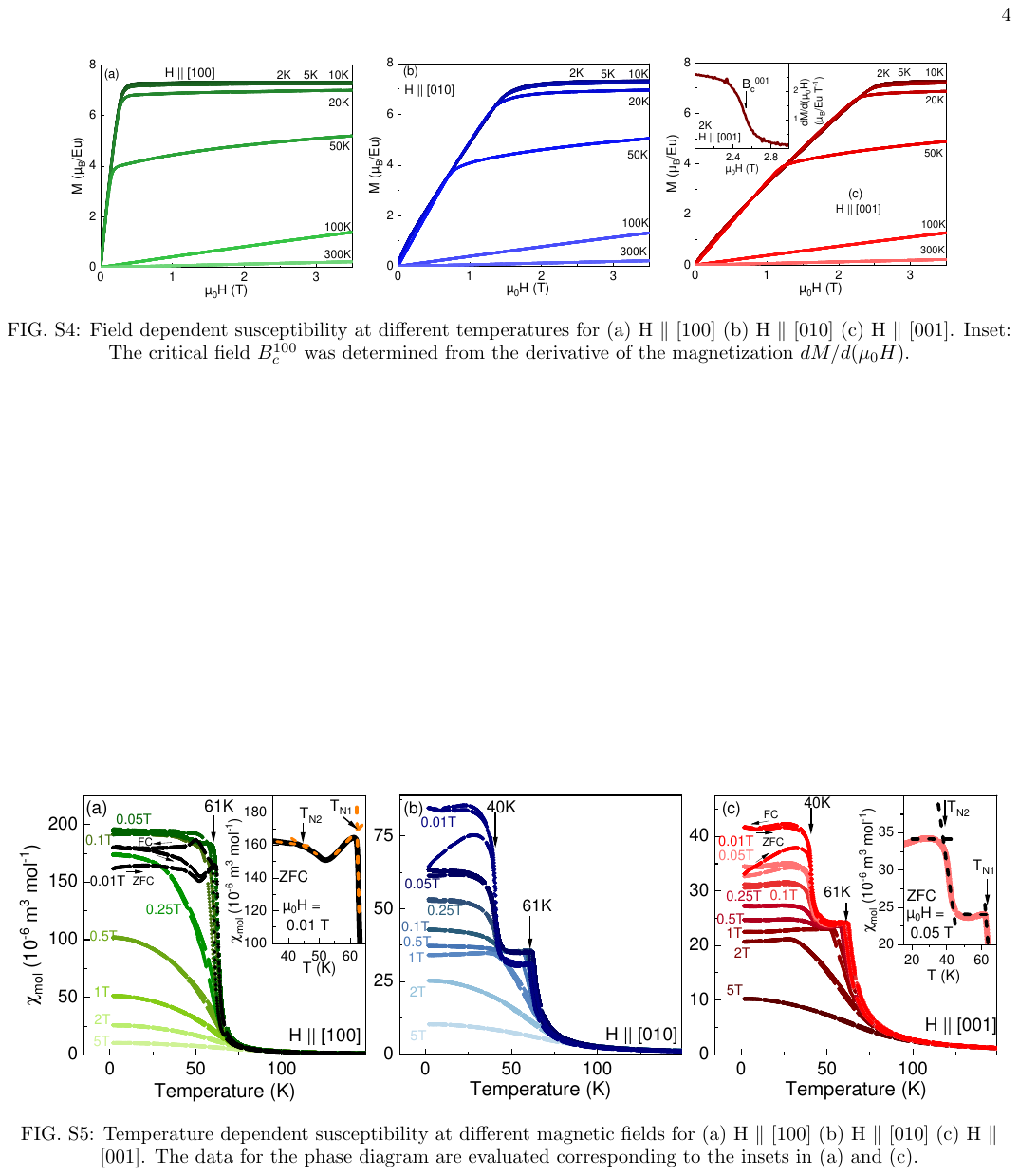}
\end{figure*}

\end{document}